\documentclass{PoS}

\usepackage{amssymb, amsthm, amsmath}
\usepackage{graphicx,graphics}
\usepackage{psfrag}
\usepackage{verbatim}
\usepackage{subfigure}
\usepackage{wrapfig}
\usepackage{amsfonts}

\title{Complex saddle points in finite-density QCD}

\ShortTitle{Complex saddle points in finite-density QCD}

\author{\speaker{Hiromichi Nishimura}\thanks{HN greatly acknowledges the support of Bielefeld University.}\\
        Bielefeld University \\
        E-mail: \email{nishimura@physik.uni-bielefeld.de}}

\author{Michael C. Ogilvie\\
     Washington University in St. Louis\\
        E-mail: \email{mco@wuphys.wustl.edu}}

\author{Kamal Pangeni\\
     Washington University in St. Louis\\
        E-mail: \email{kamalpangeni@wustl.edu}}

\abstract{We consider complex saddle points in QCD at finite temperature and density, which are constrained by symmetry under charge and complex conjugations. This approach naturally incorporates color neutrality, and the Polyakov loop and the conjugate loop at the saddle point are real but not identical.  Moreover, it can give rise to a complex mass matrix associated with the Polyakov loops, reflecting oscillatory behavior in color-charge densities.  This aspect of the phase structure appears to be sensitive to the origin of confinement, as modeled in the effective potential.}

\FullConference{9th International Workshop on Critical Point and Onset of Deconfinement - CPOD2014,\\
		17-21 November 2014\\
		ZiF (Center of Interdisciplinary Research), University of Bielefeld, Germany}

\begin{document}

\section{Background}
\label{sec:background}

The phase structure of QCD at finite density and temperature is of fundamental importance, and can be studied both experimentally and theoretically.
Nevertheless, progress has been slow, in part because of the sign problem, which afflicts both lattice simulations \cite{deForcrand:2010ys} and phenomenological models \cite{Dumitru:2005ng}.
The problem in QCD is due to the fact that the fermion determinant is complex
for typical gauge field configurations when the quark chemical potential
$\mu$ is nonzero.
In \cite{Nishimura:2014rxa}, we have shown that the consideration of complex saddle points provides
a conceptually cohesive phenomenological model of QCD at finite $T$
and $\mu$. 
Moreover, we have identified a new property of QCD at finite density, the occurrence of a disorder line, that may have observable consequences in experiment and/or lattice simulation. 
Some feature associated with the disorder line differentiate strongly
between different phenomenological models, and may thus have an impact
on our understanding of confinement.

\begin{figure}
\begin{center}
\subfigure[$\mu=0$]{\includegraphics[width=2.9in]{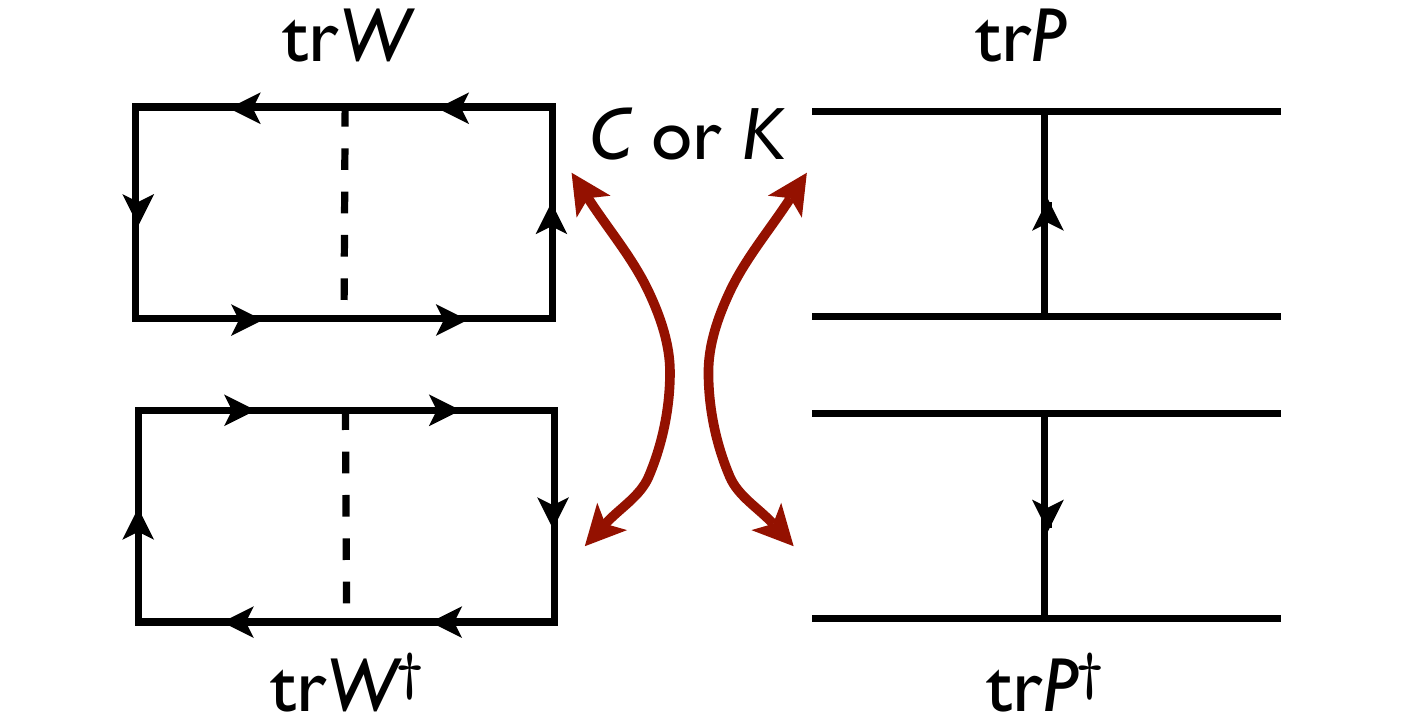}\label{fig:zero_mu}}
\subfigure[$\mu \neq 0$]{\includegraphics[width=2.9in]{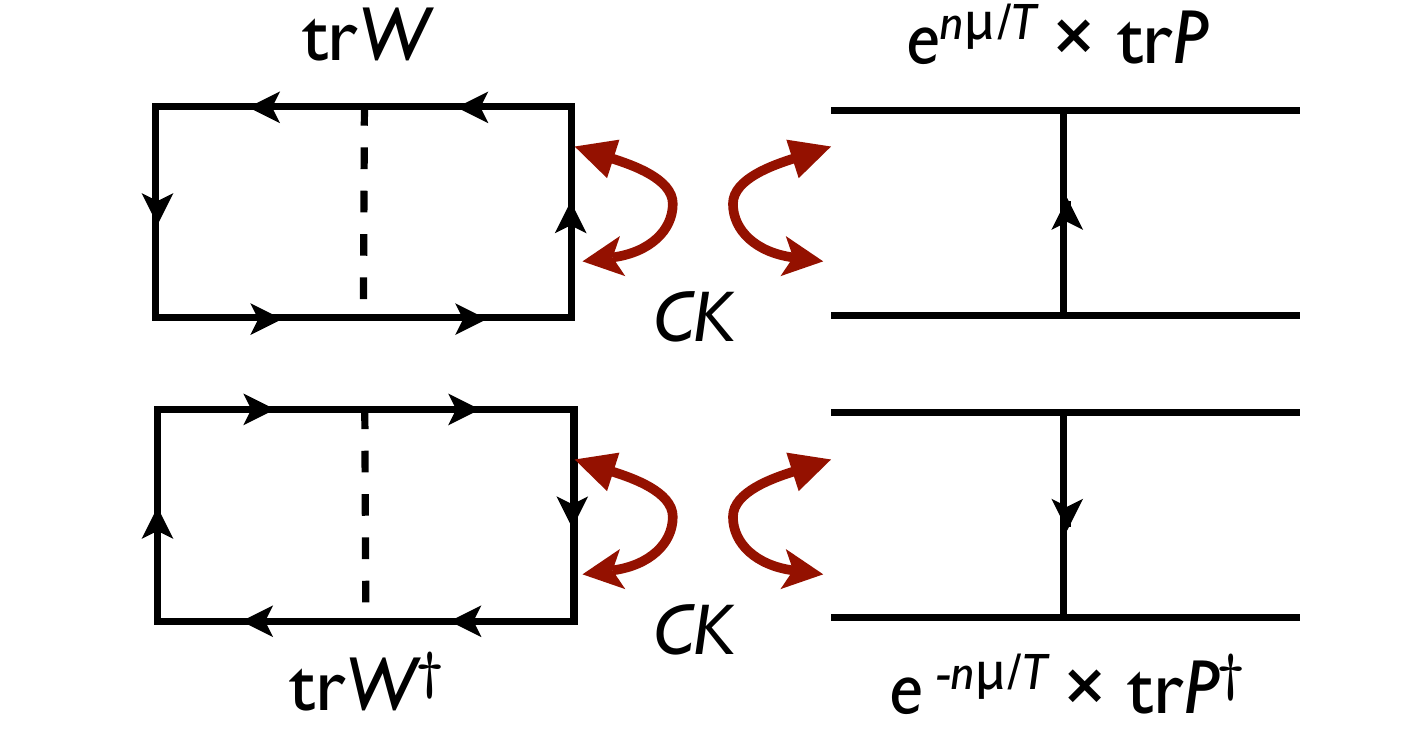}\label{fig:nonzero_mu}}
\vspace{-2mm}
\caption{The Wilson loop $\mbox{tr} W$ and the Polyakov loop $\mbox{tr}P$ under the charge conjugation $\mathcal{C}$ and the complex conjugation $\mathcal{K}$.} \label{fig1}
\end{center}
\end{figure}

We consider an $SU(N)$ gauge theory coupled to fermions in the fundamental representation. It is well-known that the Euclidean Dirac operator has complex eigenvalues when a nonzero chemical potential is introduced. This can be understood as an explicit breaking of charge conjugation symmetry $\mathcal{C}$.
The log of the fermion determinant, $\log\det\left(\mu,A\right)$, which is a function of the quark chemical potential $\mu$ and the gauge field $A$, can be formally expanded as a sum over Wilson loops with real coefficients. For a gauge theory at finite temperature, the sum includes Wilson loops that wind nontrivially around the Euclidean
timelike direction; Polyakov loops are examples of such loops. At $\mu=0$, every Wilson loop ${\rm Tr}_{F}W$
appearing in the expression for the fermion determinant is combined
with its conjugate ${\rm Tr}_{F}W^{\dagger}$ to give a real contribution
to path integral weighting. More formally, charge conjugation acts
on matrix-valued Hermitian gauge fields as
\begin{equation}
\mathcal{C}:\, A_{\mu}\rightarrow-A_{\mu}^{t},
\end{equation}
where the overall minus sign is familiar from QED, and the transpose
interchanges particle and antiparticle, \emph{e.g.}, $W^{+}$ and
$W^{-}$ in $SU(2).$ This transformation law in turn implies that
$\mathcal{C}$ exchanges the Wilson loop and the conjugate loop as shown in Fig.~\ref{fig:zero_mu},
so unbroken charge symmetry implies a real fermion determinant. 
When $\mu\ne0$, Wilson loops with nontrivial winding number $n$ in the
$x_{4}$ direction receive a weight $e^{n\beta\mu}$ while the conjugate
loop is weighted by $e^{-n\beta\mu}$ as illustrated in Fig.~\ref{fig:nonzero_mu}, and thus invariance under $\mathcal{C}$
is explicitly broken. However, there is a related antilinear symmetry
which is unbroken: ${\rm Tr}_{F}W$ transforms into itself under the
combined action of $\mathcal{CK}$, where $\mathcal{K}$ is the fundamental
antilinear operation of complex conjugation. 
Thus the theory is invariant under $\mathcal{CK}$ even in the case $\mu\ne0$. 
Note that $\mathcal{K}$ itself is also a symmetry of the theory when $\mu=0$, but it is explicitly broken when $\mu\neq 0$ for the same reason as $\mathcal{C}$. $\mathcal{CK}$ symmetry is
an example of a generalized $\mathcal{PT}$ (parity-time) symmetry
transformation \cite{Bender:1998ke}; theories with
such symmetries form special class among theories with sign problems.
Even though $\mathcal{CK}$ is a trivial transformation for each Wilson loop, it implies the well-known relation
$\det\left(-\mu,A_{\mu}\right)=\det\left(\mu,A_{\mu}\right)^{*}$
for Hermitian $A_{\mu}$, because acting $\mathcal{C}$ on the fermion determinant is equivalent to changing $\mu$ to $-\mu$ as one can see from Fig.~\ref{fig1}, and acting $\mathcal{K}$ on the determinant simply gives the complex conjugation. The advantage of using $\mathcal{CK}$ is that it is more general, leading to more insight into the sign problem and applying to bosons as well as to fermions.

For phenomenological models, the existence of $\mathcal{CK}$ symmetry
leads naturally to the consideration of complex but $\mathcal{CK}$-symmetric
saddle points.
Typically, such models require the minimization of some effective
action $\Gamma$ or effective potential $V_{\rm{eff}}$ as a function
of some set of fields. We will consider models with
effective potentials that are class functions of
the Polyakov loop $P$, depending only on the set of eigenvalues
of $P$.
$\mathcal{CK}$ symmetry will map any saddle-point
configuration $A_{\mu}^{(1)}$ into another saddle point given by
$A_{\mu}^{(2)}=-A_{\mu}^{(1)\dagger}$ with a corresponding connection
between the actions of the two configurations: $S^{(2)}=S^{(1)*}$.
However, some field configurations are themselves $\mathcal{CK}$-symmetric
in that $-A_{\mu}^{\dagger}$ is equivalent to $A_{\mu}$ under a
gauge transformation. If a saddle point is $\mathcal{CK}$ symmetric,
then its action and effective potential are necessarily real. A quick
direct proof can be given: For such a field configuration, it is easy
to prove that every Wilson loop is real and thus $\det\left(\mu,A_{\mu}\right)$
is real and positive for a $\mathcal{CK}$-symmetric field configuration.
If a single $\mathcal{CK}$-symmetric saddle point dominates the effective
potential, then the sign problem is solved, at least for a particular
phenomenological model. Such $\mathcal{CK}$-symmetric saddle points
have been seen before in finite density calculations \cite{Hands:2010vw}.

Let us consider the Polyakov loop $P$, a special kind of Wilson loop, associated with some particular field configuration that is $\mathcal{CK}$-symmetric.
We can transform to Polyakov gauge where $A_{4}$ is diagonal and time-independent, and work with the eigenvalues $\theta_{j}$ defined by
\begin{equation}
P\left(\vec{x}\right)=\mbox{diag}\left[e^{i\theta_{1}\left(\vec{x}\right)},\cdots,\, e^{i\theta_{N}\left(\vec{x}\right)}\right],
\end{equation}
where the $\theta_{j}$'s are here complex but satisfy $\sum_{j}\theta_{j}=0$. 
Because we are primarily interested in constant saddle points, we
suppress the spatial dependence hereafter. Invariance under $\mathcal{CK}$
means that the set $\left\{ -\theta_{j}^{*}\right\} $ is equivalent
to the $\left\{ \theta_{j}\right\} $ although the eigenvalues themselves
may permute. In the case of $SU(3)$, we may write this set uniquely
as
\begin{equation}
\left\{ \theta-i\psi,-\theta-i\psi,2i\psi\right\} .\label{eq:EVs_PolyakovLoop_SU3}
\end{equation}
This parametrizes the set of $\mathcal{CK}$-symmetric $SU(3)$ Polyakov
loops. 
Recalling the form of the diagonal Gell-Mann matrices
$\lambda_3$ and $\lambda_8$,
we see that
in Polyakov gauge  $\theta \ne 0$ corresponds to a nonzero
real value for $A_4^3$, while $\psi \ne 0$ corresponds to a purely
imaginary value for $A_4^8$.
Thus a $\mathcal{CK}$-symmetric saddle point requires
analytic continuation of $A_4^8$ along the imaginary axis.
Notice that both
\begin{equation}
{\rm Tr}_{F}P=e^{\psi}2\cos\theta+e^{-2\psi}
\end{equation}
 and
\begin{equation}
{\rm Tr}_{F}P^{\dagger}=e^{-\psi}2\cos\theta+e^{2\psi}
\end{equation}
are real, but they are equal only if $\psi=0$. In the usual interpretation
of the Polyakov loop expectation value, this implies that the free
energy change associated with the insertion of a fermion is different
from the free energy change associated with its antiparticle. 
It is easy to check that the trace of all powers of $P$ or $P^{\dagger}$ are all real, and thus all group characters are real as well.

The existence of complex $\mathcal{CK}$-symmetric saddle points provides
a fundamental approach to non-Abelian gauge theories that is similar
to the heuristic introduction of color chemical potentials, and naturally
ensures the system has zero color charge, \emph{i.e.}, all three charges
contribute equally \cite{Buballa:2005bv}. In the case of $SU(3)$,
extremization of the thermodynamic potential with respect to $\theta$
leads to the requirement $\left\langle n_{r}\right\rangle -\left\langle n_{g}\right\rangle =0$
where $\left\langle n_{r}\right\rangle $ is red color density, including
the contribution of gluons. Similarly, extremization of the thermodynamic
potential with respect to $\psi$ leads $\left\langle n_{r}\right\rangle +\left\langle n_{g}\right\rangle -2\left\langle n_{b}\right\rangle =0$.
Taken together, these two relations imply that $\left\langle n_{r}\right\rangle =\left\langle n_{g}\right\rangle =\left\langle n_{b}\right\rangle $.

We demand that any saddle point solution be stable to constant, real
changes in the Polyakov loop eigenvalues, corresponding for $SU(3)$
to constant real changes in $A_{4}^{3}$ and $A_{4}^{8}$. Consider
the $\left(N-1\right)\times\left(N-1\right)$ matrix $M_{ab}$, defined
in Polyakov gauge as 
\begin{equation}
M_{ab}\equiv g^{2}\frac{\partial^{2}V_{\rm{eff}}}{\partial A_{4}^{a}\partial A_{4}^{b}}.
\end{equation}
At very high temperatures and densities, the eigenvalues of this mass
matrix give the usual Debye screening masses. The stability criterion
is that the eigenvalues of $M$ must have positive real parts. At
$\mathcal{CK}$-symmetric saddle points, the eigenvalues will be either
real or part of a complex conjugate pair. In the case of $SU(3),$
the matrix $M$ may also be written in terms of derivatives with respect
to $\theta$ and $\psi$ as
\begin{equation}
M=\frac{g^{2}}{T^{2}}\left(\begin{array}{cc}
\frac{1}{4}\frac{\partial^{2}V_{\rm{eff}}}{\partial\theta^{2}} & \frac{i}{4\sqrt{3}}\frac{\partial^{2}V_{\rm{eff}}}{\partial\theta\partial\psi}\\
\frac{i}{4\sqrt{3}}\frac{\partial^{2}V_{\rm{eff}}}{\partial\theta\partial\psi} & \frac{-1}{12}\frac{\partial^{2}V_{\rm{eff}}}{\partial\psi^{2}}
\end{array}\right).\label{eq:M_mass_matrix}
\end{equation}
This stability criterion generalizes the stability criterion used
previously for color chemical potentials, which was $\partial^{2}V_{\rm{eff}}/\partial\psi^{2}<0$.
Crucially, the mass matrix $M_{ab}$ is invariant under $M^{*}=\sigma_{3}M\sigma_{3}$,
which is itself a generalized $\mathcal{PT}$ (parity-time) symmetry
transformation \cite{Bender:1998ke}. It is easy
to see that this relation implies that $M_{ab}$ has either two real
eigenvalues or a complex eigenvalue pair. In either case, the real
part of the eigenvalues must be positive for stability. In the case
where there are two real eigenvalues, we will denote by $\kappa_{1}$
and $\kappa_{2}$ the two positive numbers such that $\kappa_{1}^{2}$
and $\kappa_{2}^{2}$ are the eigenvalues of the mass matrix $M_{ab}$.
If $M_{ab}$ has two complex eigenvalues, we define two positive real
numbers $\kappa_{R}$ and $\kappa_{I}$ such that $\left(\kappa_{R}\pm i\kappa_{I}\right)^{2}$
are the conjugate eigenvalues of $M_{ab}.$ The border separating
the region $\kappa_{I}\ne0$ from the region $\kappa_{I}=0$ is known
as the disorder line.
In this case, it separates the region where the color density correlation
function decays exponentially in the usual way from the region where
a sinusoidal modulation is imposed on that decay.

\section{Models}

We now consider a class of phenomenological models that combines the one-loop result with the effects of confinement for the case of $SU(3)$ gauge bosons and two flavors of quarks at finite temperature and density.
The model is described by an effective potential which is the sum of three terms: 
\begin{equation}
V_{\rm{eff}}(P)=V_{g}(P)+V_{f}(P)+V_{d}(P).
\end{equation}
The potential term $V_{g}(P)$ is the one-loop effective potential for gluons.  The potential term $V_{f}\left(P\right)$ contains all quark effects, including the one-loop expression. The potential term $V_{d}\left(P\right)$ represents confinement effects. 
We will consider three different forms for $V_{f}\left(P\right)$ and two different forms for $V_{d}\left(P\right)$
for a total of six different models. The formulas and parameters we use for these models can be found in \cite{Nishimura:2014rxa}.

The potential term $V_{d}(P)$ is taken to respect center symmetry and acts to favor the confined phase at
low temperature and density \cite{Meisinger:2001cq,Myers:2007vc}. The gauge contribution $V_{g}(P)$
favors the deconfined phase, and in the pure gauge theory ($N_{f}=0$)
the deconfinement transition arises out of the competition between
$V_{g}(P)$ and $V_{d}(P)$. 
The parameters of $V_{d}(P)$ are set to reproduce the deconfinement temperature of the pure gauge theory, known from lattice simulations to occur at $T_{d}\approx270\,\mbox{MeV}$. 
The specific forms used here are Model A and Model B of \cite{Meisinger:2001cq}, which can be written as
\begin{eqnarray}
V_{d}^{A}\left(P\right)
&=&\sum_{j,k=1}^{N}(1-\frac{1}{N}\delta_{jk})\frac{M_{A}^{2}}{2\beta^{2}}B_{2}\left(\frac{\Delta\theta_{jk}}{2\pi}\right)
=
\frac{M_{A}^{2}T^{2}\left((2\pi-3\theta)^{2}-27\psi^{2}\right)}{6\pi^{2}},
\\
V_{d}^{B}(P)&=&-\frac{T}{R^{3}}\log\left[\prod_{j<k}\sin^{2}\left(\frac{\theta_{j}-\theta_{k}}{2}\right)\right] 
= -\frac{T}{R^{3}}\log\left[\frac{1}{4}\left\{ \cos\theta-\cosh\left(3\psi\right)\right\} ^{2}\sin^{2}\theta\right],
\end{eqnarray}
where we have used the parametrization in Eq.~(\ref{eq:EVs_PolyakovLoop_SU3}).
$\Delta\theta_{jk}=\left|\theta_{j}-\theta_{k}\right|$ are
the adjoint Polyakov loop eigenvalues and $B_{2}$ is the second Bernoulli
polynomial. The expression for Model A gives a simple quartic polynomial in the
Polyakov loop eigenvalues for $V_{g}\left(P\right)+V_{d}^{A}\left(P\right)$
and thus can be thought of as a form of Landau-Ginsburg potential
for the Polyakov loop eigenvalues. 
The parameter $M_{A}$ controls the location of the deconfinement
transition in the pure gauge theory, and is set to $596\,\mbox{MeV}$.
On the other hand, the form for Model B is motivated by Haar measure, representing
a determinant term that tries to keep a space-time volume of order
$\beta R^{3}$ confined. 
In order to reproduce the correct deconfinement temperature for the
pure gauge theory, $R$ must be set to $R=1.0028$ fm. 
At low temperatures, the potential $V_d(P)$ dominates the pure gauge theory effective
potential. The variable $\psi$ is zero, and $V_{d}\left(P\right)$
is minimized when $\theta=2\pi/3$. For this value of $\theta$, the
eigenvalues of $P$ are uniformly spaced around the unit circle, respecting
center symmetry, and $\mathrm{Tr}_{F}P=\mathrm{Tr}_{F}P^{\dagger}=0$.
As the temperature increases, $V_{g}\left(P\right)$ becomes relevant,
and gives rise to the deconfined phase where center symmetry is spontaneously
broken. The addition of light fundamental quarks via $V_{f}(P)$ explicitly
breaks center symmetry. For all nonzero temperatures, center symmetry
is broken and $\left\langle \mathrm{Tr}_{F}P\right\rangle \ne0$.
However, a remnant of the deconfinement transition remains in the
form of a rapid crossover from smaller value of $\mathrm{Tr}_{F}P$
to larger ones as $T$ and $\mu$ are varied. 

Although $V_{d}^{A}$ and $V_{d}^{B}$ appear to be very different,
and are motivated in different ways, they are actually closely related.
The confining potentials $V_{d}$ can also be written as
\begin{eqnarray}
V_{d}^{A}&=&\frac{M_{A}^{2}T^{2}}{2\pi^{2}}\sum_{n=1}^{\infty}\frac{1}{n^{2}}{\rm Tr}_{A}P^{n},
\\
V_{d}^{B}&=&\frac{T}{R^{3}}\sum_{n=1}^{\infty}\frac{1}{n}{\rm Tr}_{A}P^{n},
\end{eqnarray}
up to some irrelevant constant for the latter case.  
Using ${\rm Tr}_{A}P={\rm Tr}_{F}P^{n}{\rm Tr}_{F}P^{\dagger n}-1$,
it is easy to prove that minimizing either $V_{d}^{A}$ or $V_{d}^{B}$
yields a confining phase where ${\rm Tr}_{F}P^{n}=0$ for all $n\ne0\,\mathrm{mod}(N)$. 

We consider three different cases of quarks with mass $m$. The first is heavy quarks,
with a fixed mass of $2$ GeV. In this model, the quarks are essentially
irrelevant for the deconfinement transition, which occurs at essentially
the same temperature as if no quarks were present at all. The effect
of spontaneous chiral symmetry breaking is not included, as it would
only contribute a small amount to the quark mass. This case is in
some sense the simplest, and perhaps would be the easiest for which
to obtain reliable simulation results. The second case considered
is massless quarks, where the fermion mass
is set equal to zero by hand. This case cannot be easily simulated using lattice
methods, because it ignores chiral symmetry breaking effects which
do occur in lattice simulations. It is thus useful only for sufficiently
large values of $T$ and $\mu$ such that chiral symmetry is essentially
restored. Our most realistic treatment of quarks uses a Nambu-Jona
Lasinio four-fermion interaction to model chiral symmetry breaking
effects, so these models are of Polyakov-Nambu-Jona Lasinio (PNJL)
type \cite{Fukushima:2003fw}.

\section{Disorder lines in phenomenological models of QCD}

We now show some results of the disorder lines, which are the borders separating the region $\kappa_I \neq 0$ from the region $\kappa_I=0$: see the last paragraph in Sec.~\ref{sec:background}. More results and discussions of the disorder lines and the Polyakov loops at the $\mathcal{CK}$-symmetric saddle point can be found in \cite{Nishimura:2014rxa}.

\begin{figure}
\begin{center}
\subfigure[Model A]{\includegraphics[width=2.9in]{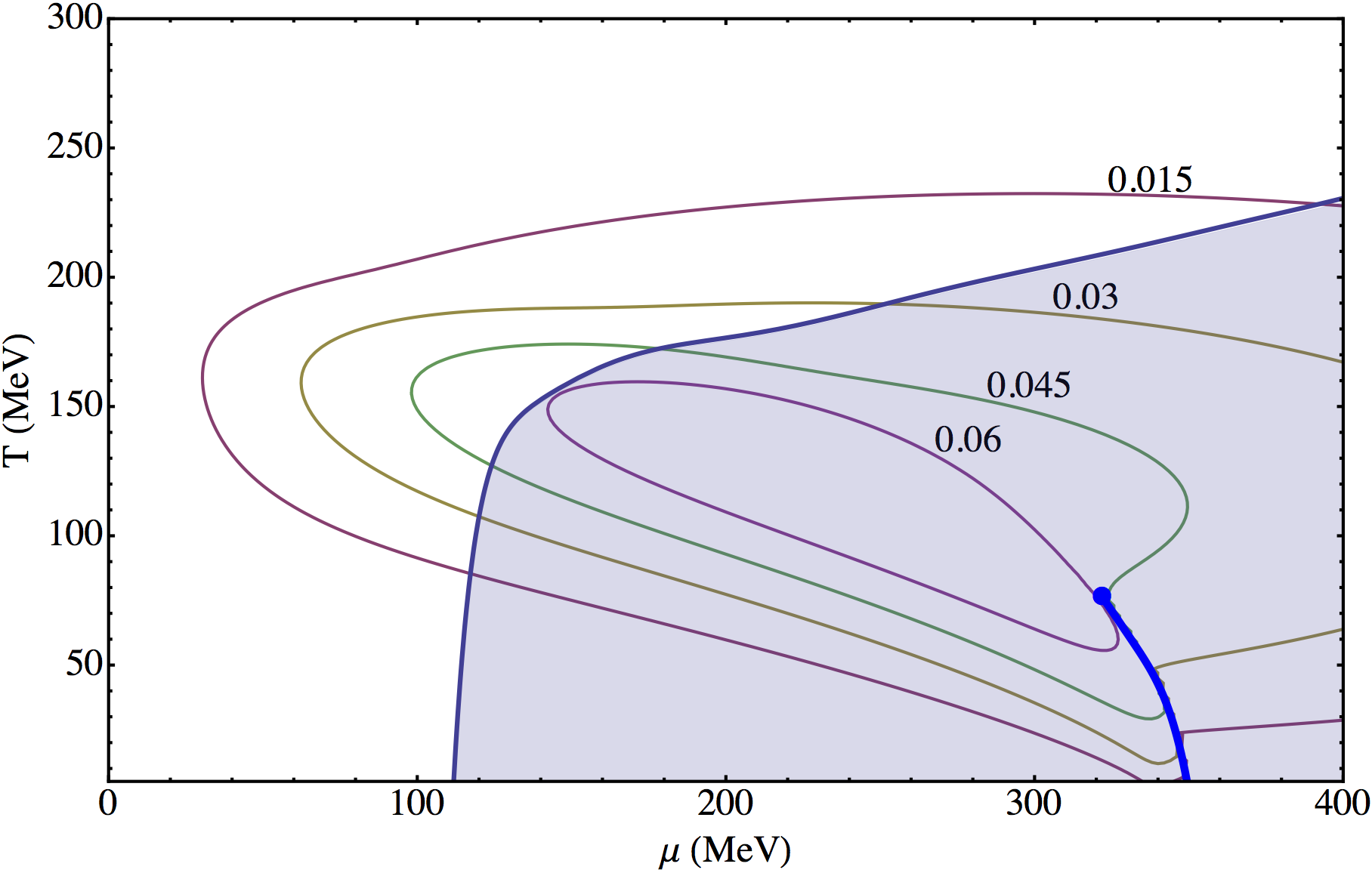}\label{fig:PNJL_A_Psi_DisorderLine_CriticalLine}}
\subfigure[Model B]{\includegraphics[width=2.9in]{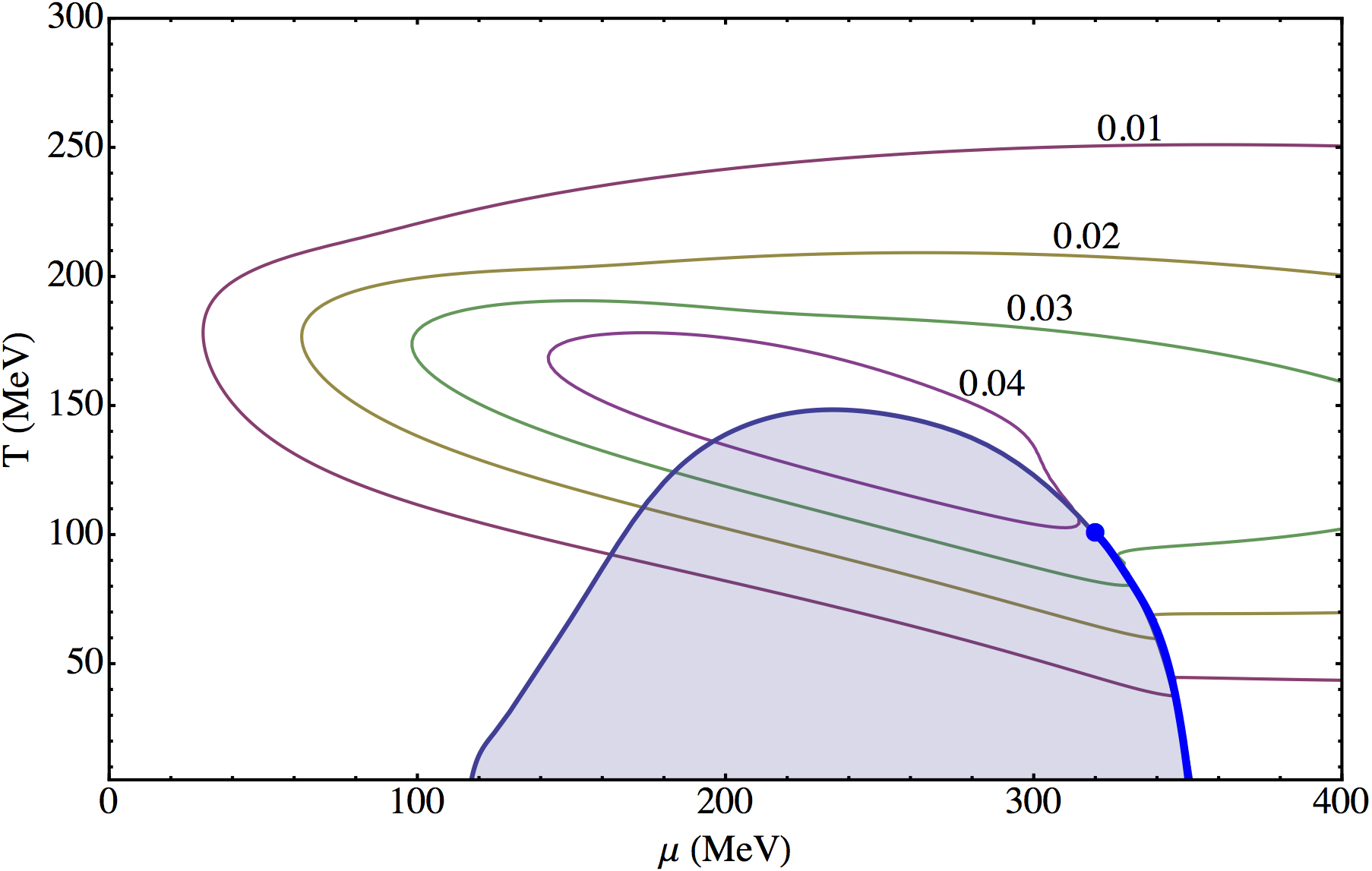}\label{fig:PNJL_B_Psi_DisorderLine_CriticalLine}}
\vspace{-2mm}
\caption{Contour plot of $\psi$ in the $\mu-T$ plane for the PNJL model. The region where $\kappa_{I}\ne0$ is shaded. The critical
line and its endpoint are also shown.} \label{fig2}
\end{center}
\end{figure}

\begin{figure}
\begin{center}
\subfigure[Model A]{\includegraphics[width=2.9in]{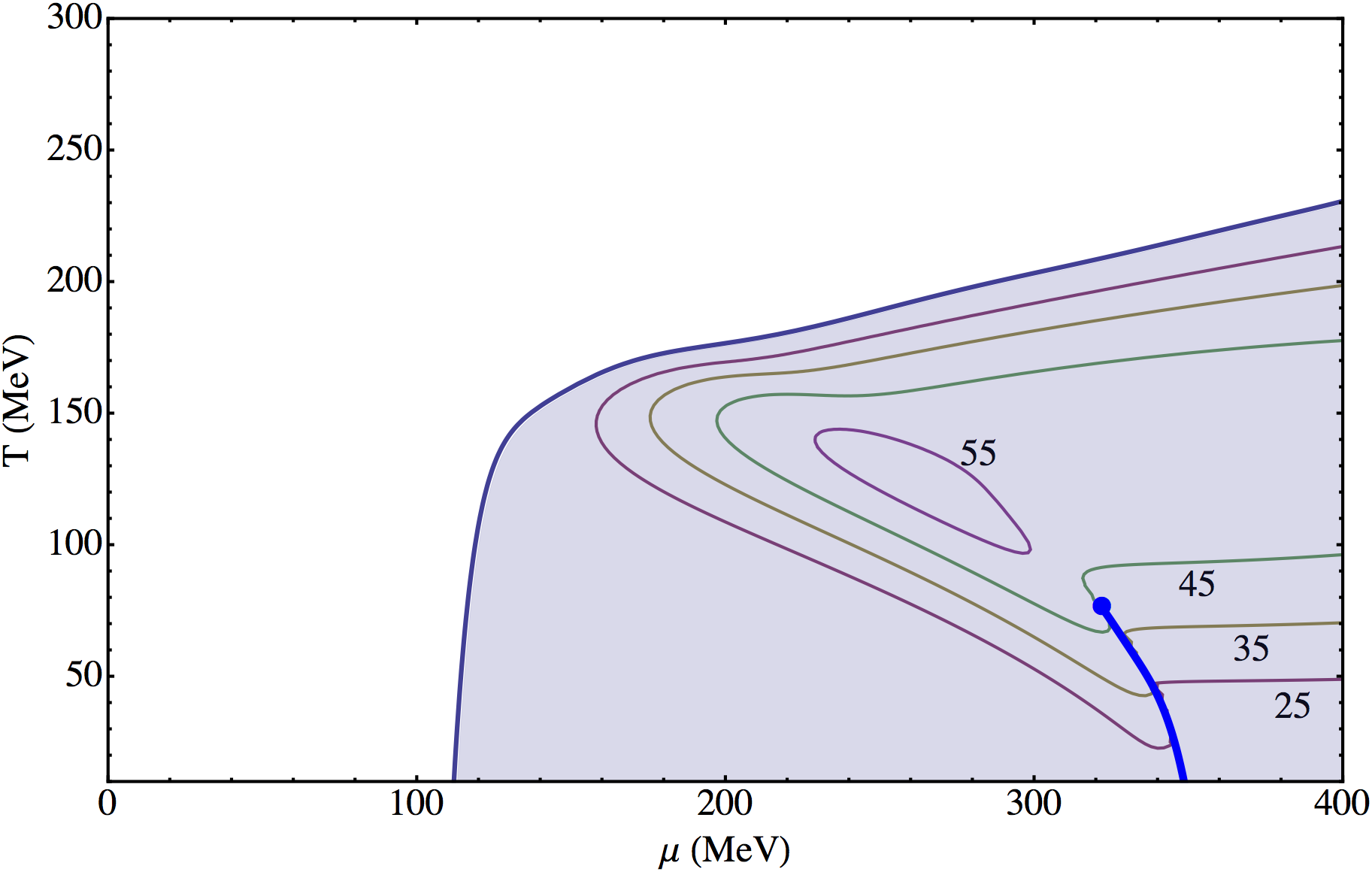}\label{fig:PNJL_A_ImaginaryMass_DisorderLine_CriticalLine}}
\subfigure[Model B]{\includegraphics[width=2.9in]{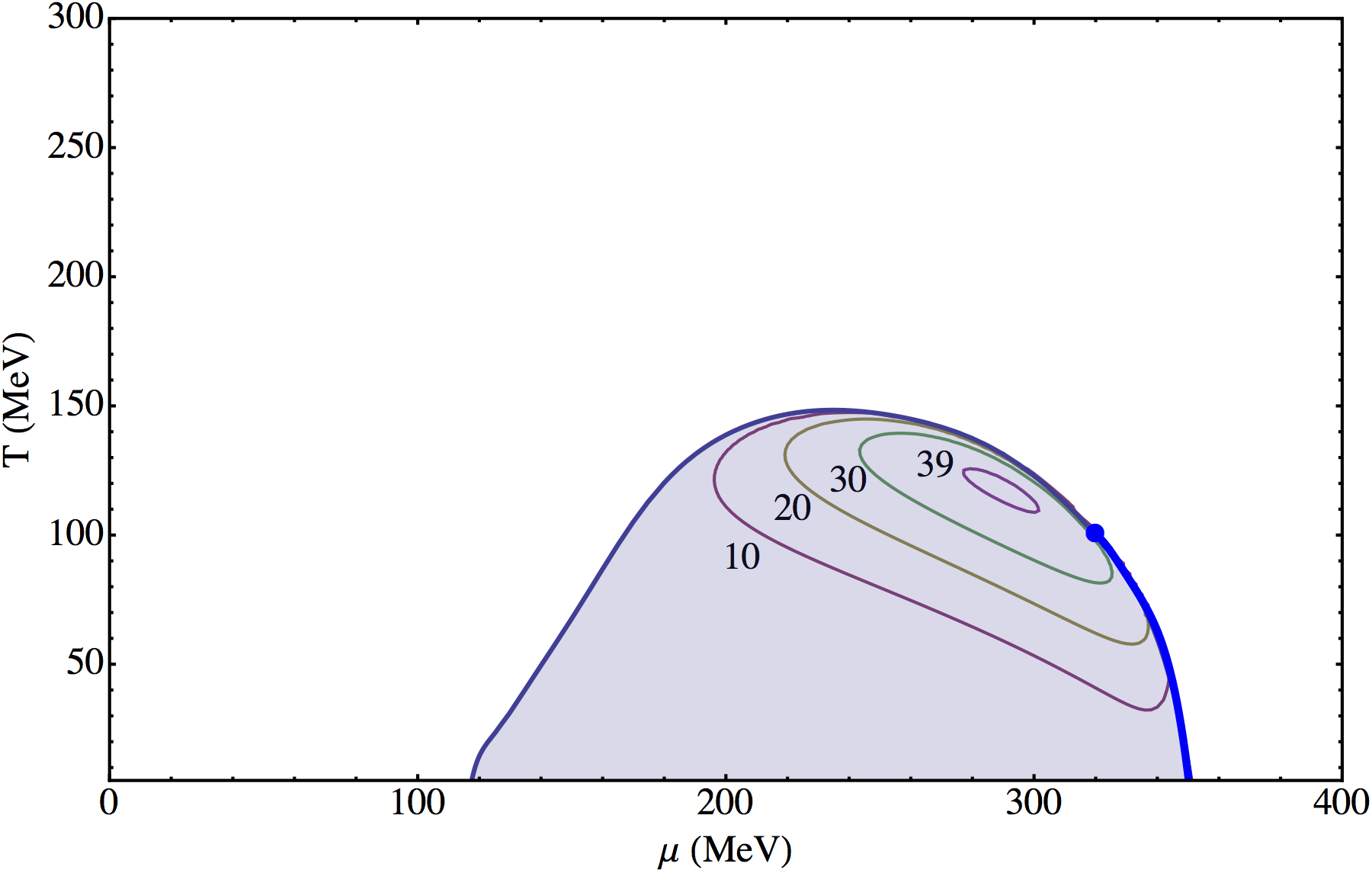}\label{fig:PNJL_B_ImaginaryMass_DisorderLine_CriticalLine}}
\vspace{-2mm}
\caption{Contour plot of $\kappa_{I}$ in the $\mu-T$ plane for the PNJL model. Contours are given in MeV, with $\alpha_{s}$ set to one. } \label{fig3}
\end{center}
\end{figure}

Figure \ref{fig2} shows contour lines for $\psi$ in the $\mu-T$ plane along with the region where
$\kappa_{I}\ne0$ as well as the critical line for the PNJL model with Model A and Model B. 
In both models, the shaded regions where $\kappa_I \neq 0$ cover a large portion of the phase diagram.
 While the critical line does not depend much on the models, there is a striking difference in the disorder line. The critical line lies completely within the region $\kappa_{I}\ne0$ for Model A while it appeases to be a smooth continuation of the critical line out of the critical end point for Model B. 
In both models, $\psi$ is very small and a jump in $\psi$ is visible as the critical line is crossed. 
Figure \ref{fig3} shows similar plots as Fig.~\ref{fig2} but with contour lines for $\kappa_I$ in the $\mu-T$ plane. $\kappa_I$ jumps at the critical line only for Model A, because the critical line lies on the disorder line for Model B. 
Comparison of the two figures shows that the peak in $\psi$ occurs at a lower value of $\mu$ than the
peak in $\kappa_{I}$, with the peak in $\psi$ occurring near ($\mu=200$
MeV, $T=110$ MeV) for Model A and ($\mu=250$ MeV, $T=140$ MeV) for Model B. 

\begin{figure}
\begin{center}
\subfigure[Massless quarks]{\includegraphics[width=2.9in]{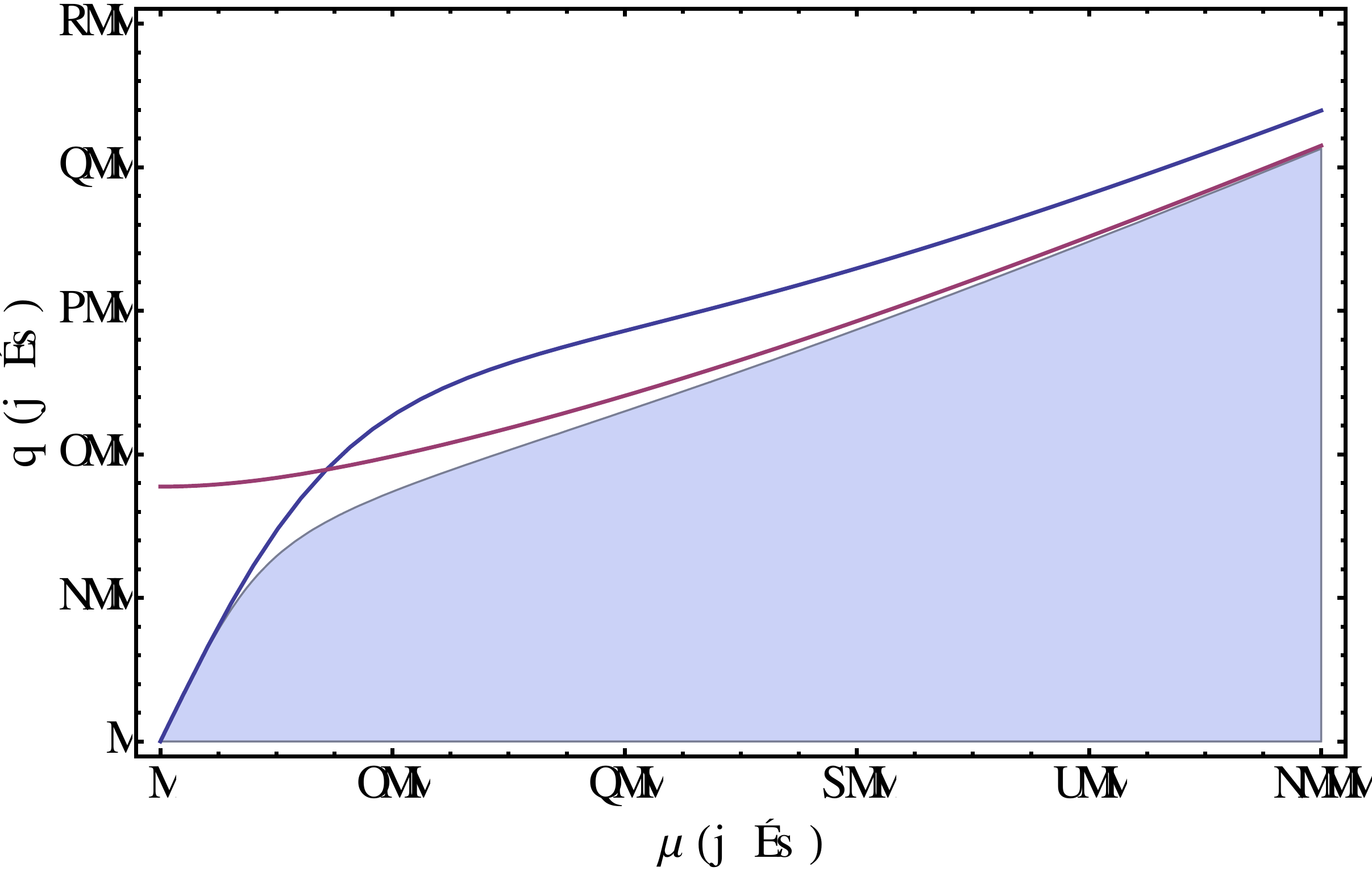}\label{fig:TmuLine3}}
\subfigure[Massive quarks ($2$ GeV)]{\includegraphics[width=2.9in]{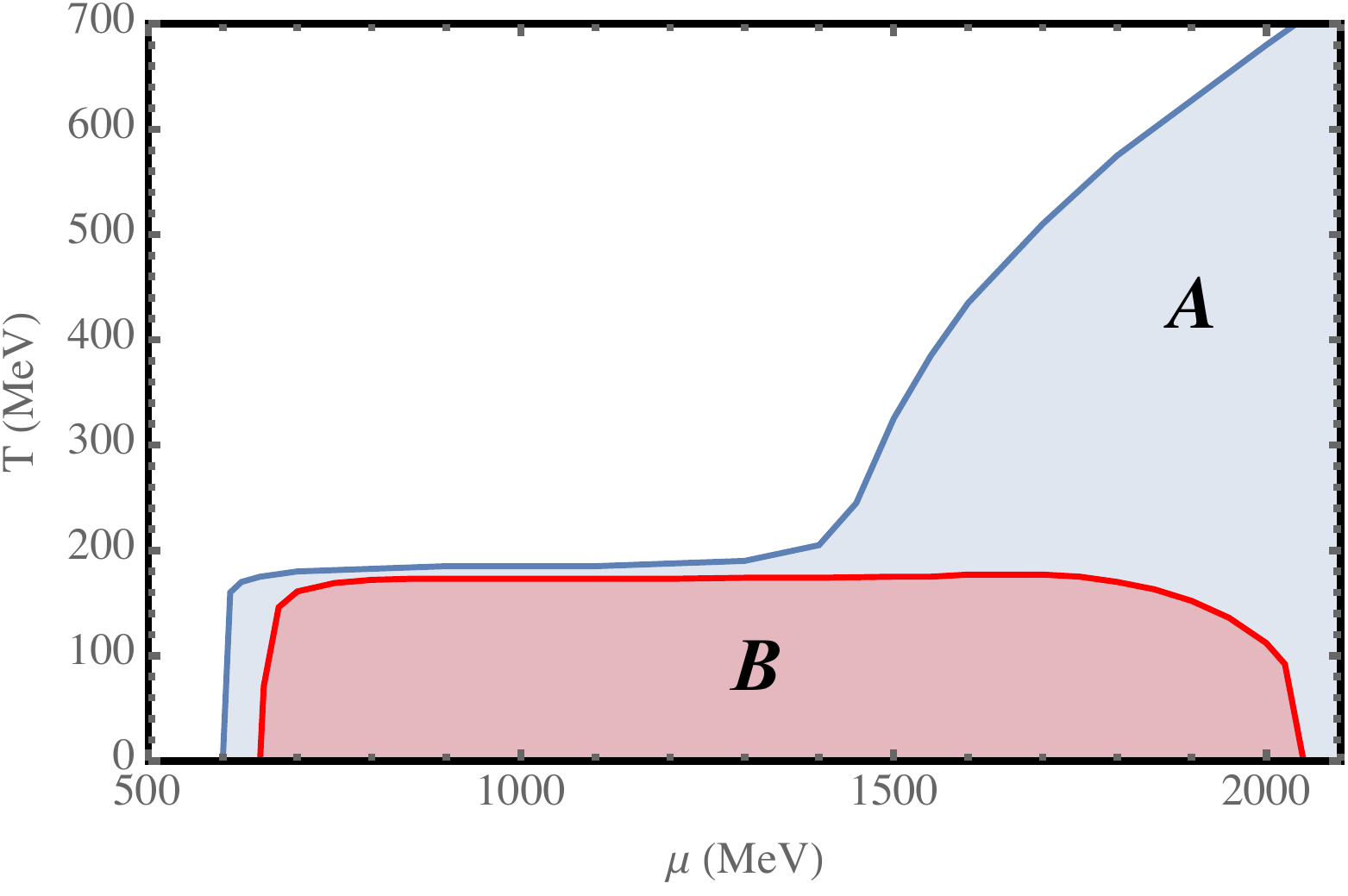}\label{fig:Comparison-AB-hq}}
\vspace{-2mm}
\caption{A comparison of the regions where $\kappa_{I}\ne0$. There is no disorder line for the case of massless quarks with Model B. High-T and low-T approximations are also shown for massless quarks.} \label{fig4}
\end{center}
\end{figure}

The behavior of the disorder line for large $T$ and $\mu$ for Model A is known analytically \cite{Nishimura:2014rxa}:
\begin{equation}
T=\frac{2\mu}{\sqrt{3}\pi}.
\end{equation}
This behavior is generic to Model A when $T,\mu\gg m$, as one can see in Figs.~\ref{fig2}-\ref{fig4}.
On the other hand, there is no disorder line in the limit $T,\mu\gg m$ for Model B.
This is consistent with the fact that there is no disorder line in the entire phase diagram for massless quarks with Model B. 
This is the only case we have considered where there is no disorder line. 
Nevertheless, as shown in \cite{Nishimura:2014rxa}, $\psi$ is nonzero even in this case, with
a peak value near the same point as in the PNJL model with Model B. 
For heavy quarks, both models have the disorder line as can be seen in Fig.~\ref{fig:Comparison-AB-hq}.
Their shape is very similar for smaller values of $\mu$, suggesting that some universal behavior occurs in this region. However, the behavior is very different in the region where both $T$ and $\mu$ are becoming large. Model A shows a continuation of the disorder line that follows the behavior for massless quarks, while for Model B the disorder line covers a finite region in $\mu-T$ space. The overall shape of the disorder line is similar to that found in the PNJL models, but of course shifted to a much larger value $\mu$.

\section{Conclusions}

As we have shown, the sign problem in QCD at finite density makes
it very desirable to extend real fields into the complex plane. This
extension is certainly necessary for steepest descents methods to
yield correct results.
Complex saddle points lead naturally to $\left\langle TrP\right\rangle \ne\left\langle TrP^{\dagger}\right\rangle $,
a result that is much more difficult to obtain when fields are restricted to the real axis. 
The nature of these saddle points are restricted
by $\mathcal{CK}$ symmetry. The case of a single dominant saddle
point is particularly tractable in theoretical analysis. In the class
of models we have examined, the saddle point is not far from the real
axis, as indicated by the small values of $\psi$ and corresponding
small differences between $\left\langle TrP\right\rangle $ and $\left\langle TrP^{\dagger}\right\rangle $.
This is good news for lattice simulation efforts, as it suggests only
a modest excursion into the complex plane is needed. The small value
of $\psi$ also indicates a small difference for thermodynamic quantities
such as pressure and internal energy between our work and previous
work on phenomenological models where only real fields were used.
For all six cases studied here, the maximum value of $\psi$ occurs
in the region where quark degrees of freedom are ``turning on,''
as indicated by crossover or critical behavior. In our previous work
on Model A for massless quarks \cite{Nishimura:2014rxa}, we were
able to show analytically how $\psi\ne0$ can arise from the interplay
of confinement and deconfinement when $\mu\ne0$. For the two PNJL models, it is striking that
the largest values of $\psi$ occur near the critical end point. These
predictions can be checked in lattice simulations by the direct measurement
of $\left\langle TrP\right\rangle $ and $\left\langle TrP^{\dagger}\right\rangle $
once sufficiently effective simulation algorithms are developed.

In all six cases studied, $\psi\ne0$ leads to two different eigenvalues
for the $A_{4}$ mass matrix. In five of the six cases studied, a
disorder line is found. This disorder line marks the boundary of the
region where the real parts of the mass matrix eigenvalues become
degenerate as the eigenvalues form a complex conjugate pair. In the
PNJL models, the disorder line is closely associated with the critical
line. Inside the region bounded by the disorder line, the complex
conjugate pairs gives rise to color charge density oscillations. Patel
has developed a scenario in which such oscillations might be observed
experimentally \cite{Patel:2011dp}.
Our results indicate that the oscillations may have too large a wavelength to be directly observable in experiment, although estimates based on phenomenological models should be applied cautiously. The mass matrix eigenvalues are
in principle accessible in lattice simulations via the measurement
of Polyakov loop correlation functions. A direct determination of
$\kappa_{I}$ may be difficult, but the disorder line itself could be determined from the merging of the values of $Re\left(\kappa_{1}\right)$ with $Re\left(\kappa_{2}\right)$. 

While the behavior of the Polyakov loop and the chiral condensate, as determined by
lattice simulations, do not strongly differentiate between the two
confining potential terms, Model A and Model B, the corresponding
two-point correlation functions do. The most physically relevant case
of PNJL models show both common features as well as clear differences
in the behavior of the disorder line between Model A and Model B.
In both cases, the maximum value of $\kappa_{I}$ occurs slightly
above and to the left of the critical end point in the $\mu-T$ plane,
in the vicinity of the region where the ratio $Tr_{F}P^{\dagger}/Tr_{F}P$
is largest. In Model A, the critical line is contained within the
boundary of the disorder line, but in Model B the disorder line appears
to come out of the critical end point as a continuation of the critical
line, a common behavior for disorder lines. Furthermore, in Model
A the disorder line continues diagonally in the $\mu-T$ plane for
large $\mu$ and $T$, but for Model B, the line bends over into the
critical line. With Model A there is thus a possibility that the effects
of the disorder line might be visible in the results of the Compressed
Baryonic Matter (CBM) experiment at FAIR. The disorder line also strongly
differentiates between Model A and Model B in the case of heavy quarks,
so lattice simulations of either light or heavy quarks that can locate
the disorder line have the potential to discriminate between the two
models.

\end{document}